\documentclass[prd,showpacs,amsmath,amssymb,superscriptaddress,preprintnumbers,nofootinbib]{revtex4}

\usepackage{amssymb}
\usepackage{amsmath}
\usepackage{graphicx}

\begin{document}

\newcommand{\nablab}{{\mathop {\rule{0pt}{0pt}{\nabla}}\limits^{\bot}}\rule{0pt}{0pt}}

\title{Isotropic cosmological model with aetherically active axionic dark matter}

\author{Alexander B. Balakin}
\email{Alexander.Balakin@kpfu.ru} \affiliation{Department of
General Relativity and Gravitation, Institute of Physics, Kazan
Federal University, Kremlevskaya str. 16a, Kazan 420008, Russia}

\author{Amir F. Shakirzyanov}
\email{shamirf@mail.ru} \affiliation{Department of
General Relativity and Gravitation, Institute of Physics, Kazan
Federal University, Kremlevskaya str. 16a, Kazan 420008, Russia}

\date{\today}

\begin{abstract}
In the framework of the extended Einstein-aether-axion theory we study the model of a two-level aetheric control over the evolution of a spatially isotropic homogeneous Universe filled with axionic dark matter. Two guiding functions are introduced, which depend on the expansion scalar of the aether flow, equal to the tripled Hubble function. The guiding function of the first type enters the aetheric effective metric, which modifies the kinetic term of the axionic system; the guiding function of the second type predetermines the structure of the potential of the axion field. We obtained new exact solutions of the total set of master equations of the model (with and without cosmological constant), and studied in detail four analytically solvable submodels, for which both guiding functions are reconstructed and illustrations of their behavior are presented.
\end{abstract}
\maketitle


\section{Introduction}

A century ago Alexander Friedmann formulated the prediction that our Universe expands, and this event predetermined all further development of cosmology and space sciences.
While remaining within this general concept, modern cosmology focuses on describing the details of this expansion, in particular, on the rate of expansion at different epochs.
New sensational results of observations obtained in the last decade have become the basis for restructuring our ideas about the history of the early Universe. Discovery of the gravitational radiation was the first important event, that made theorists think about the validity of previous ideas. Indeed, in 2015 the first observation of the gravitational waves from the black hole merger \cite{GW2015} put researchers in a dilemma. In this event the masses of the colliding black holes were predicted to be of 36 and 29 $M_{(\rm Sun)}$, while mass values in the range 2.5{-}10 $M_{(\rm Sun)}$ predicted by the theory of stellar collapse, seemed to be reasonable. Then the gravitational wave event indicated as GW trigger S190521g (GW 190521) \cite{GW2019}, has shown that the black holes with the masses 85 and 66  $M_{(\rm Sun)}$ collided;  the general consensus is that the mass of at least one of these black holes lies in a mass range that excludes its birth through the collapse of a star. The discovery of black hole with the so-called intermediate mass 91.000 $M_{(\rm Sun)}$ \cite{2022}, the existence of which can not be explained by the existing theories, completed the formulation of the dilemma: either it is necessary to abandon this interpretation, or admit that there is a new unknown mechanism for the formation of black holes. Fortunately, the second trend has triumphed and now theorists are actively involved in adequate extension of the models of the birth of black holes.
Another amazing story is connected with observations on the newest James Webb Space Telescope (JWST). New observational data concern, in particular, the discovery of an extremely magnified monster star, estimations of the masses of warm dark matter particles and of the axion dark matter particles \cite{JWST1}; the abundance of carbon-containing molecules \cite{JWST2}. But the most important event, from our point of view is the discovery of enormous distant galaxies that should not exist, if one follows the standard model of the early Universe evolution. To be brief, the galaxies found in the JWST images \cite{JWST3} appeared shockingly big, and the stars in them too old, and these findings are in conflict with existing models. In other words, rapid development is predicted for the theory of the evolution of the early Universe over the next few years, and modifications of the cosmological models are highly welcome.

At the moment, the most adequate picture of the world contains an early  era of inflation, epochs of domination of radiation and matter, and a late-time era of accelerated expansion. The theorist's dream is to unify the entire history of the Universe within the framework of one cosmological model (see, e.g., \cite{1,2,3,4,5,6,7}). The main obstacle to solving this problem is the difficulty of finding a unified equation of state for cosmic substrates that determine the rate of evolution of the Universe in the corresponding epoch. One of the attempts was the search for time depending parameters of the equation of state, and the introduction of a cosmological term depending on time. However, such attempts were considered unsuccessful because cosmological time is not an invariant, and therefore such equations of state are associated with the loss of covariance of the theory. A similar problem arises, when one tries to define the equation of state in terms of the redshift value $Z$, or equivalently, via the scale factor $a(t)$.

We follow another logic. We admit that the parameters of the equation of state depend on the set of scalars, which are formed on the basis of fundamental fields inherent in the cosmological model under consideration. To be more precise, we take the unit timelike vector field $U^j$ associated with the velocity four-vector of the dynamic aether \cite{J1,J2,J3,J4} and consider the invariants obtained in the course of decomposition of its covariant derivative $\nabla_k U^j$. In other words, we use four differential invariants (the expansion scalar of the aether flow,  $\Theta {=} \nabla_k U^k$, the squares of the acceleration four-vector, of the shear and vorticity tensors, $a^2$, $\sigma^2$, $\omega^2$, respectively), as the arguments of the parameters included into the equations of state. This means that we follow the paradigm of an aetheric control over the evolution of physical systems (see, e.g., \cite{B1,B2,B3,B4,B5}). We have to emphasize that depending on the spacetime symmetry of the model a part of the listed arguments can disappear. For instance, for the static spherically symmetric model we obtain that $\Theta {=}0$, $\sigma^2{=}0$, $\omega^2 {=}0$, and we construct the guiding functions using $a^2$ only. For the G\"odel spacetime the only $\omega^2$ is non-vanishing. For the spacetime with planar gravitational waves we have to work with two non-vanishing scalars: $\Theta$ and $\sigma^2$. Spatially isotropic homogeneous cosmological models are unique in this sense, since for them only the scalar $\Theta$ is non-vanishing, and this scalar coincides with the tripled Hubble function $\Theta {=} 3 H(t)$. In this context the function $H(t)$ can be chosen as an appropriate argument of the guiding parameters of such cosmological models, unifying the paradigm of the aetheric control over the physical systems evolution, on the one hand, and the physical interpretation of the theory predictions, on the other hand.  Since the function $H$ has the dimensionality of inverse time (we consider the units with $c{=}1$), this quantity is often used to determine a specific time scale in a corresponding cosmological epoch.

In this paper we work within the Einstein-aether-axion model on the Friedmann-Lema\^itre-Robertson-Walker  spacetime platform, and consider the interaction of the gravitational field, pseudoscalar (axion) field $\phi$, and unit timelike vector field $U^j$. Two guiding functions depending on the scalar $\Theta$ are introduced into the Lagrangian. The guiding function of the first type, ${\cal A}(\Theta)$ enters the so-called aetheric effective metric $G^{mn} {=} g^{mn} {+} {\cal A} U^m U^n$ (see \cite{Gcos} for history, mathematical details and motives); it modifies the kinetic term associated with the axion field, and thus it controls the evolution of the kinetic energy of the axionic dark matter in the Universe (see, e.g., \cite{Ax1}-\cite{Ax3}, which present the history of axions, and \cite{ADM1} - \cite{ADM3}, where the problem of axions in cosmology are discussed in various aspects). The guiding function of the second type, $\Phi_*(\Theta)$ enters the potential of the axion field, $V\left(\phi, \Phi_* \right)$, thus performing control over the evolution of the potential energy of the axionic dark matter.
The set of master equations of the model is solved in quadratures and partially in the analytic form; the corresponding functions ${\cal A}(\Theta)$ and $\Phi_*(\Theta)$ are reconstructed.

The paper is organized as follows. Section~\ref{sec2} contains the description of the mathematical formalism. In Section~\ref{sec3}  we analyze the key equations of the model for the spatially isotropic homogeneous cosmological model and discuss the obtained solutions. Section~\ref{sec4} contains discussion and conclusions.

\section{The formalism of the extended Einstein-aether-axion theory}\label{sec2}

\subsection{The extended action functional and auxiliary quantities}

The extended Einstein-aether-axion theory is formulated on the base of the following action functional:
$$
-S_{(\rm total)} = \int d^4x \sqrt{-g} \left\{ \frac{1}{2\kappa}\left[R{+}2\Lambda {+} \lambda\left(g_{mn}U^mU^n {-} 1 \right) {+}  {\cal K}^{ab}_{\ \ mn} \nabla_a U^m  \nabla_b U^n \right]
+ \right.
$$
\begin{equation}
\left.
+ \frac12 \Psi^2_0   \left[V(\phi, \Phi_*) {-} G^{mn} \nabla_m \phi \nabla_n \phi \right]     \right\} \,.
\label{0}
\end{equation}
In this formula the standard elements of this theory appear, such as the determinant of the spacetime metric $g$, the Ricci scalar $R$, the cosmological constant $\Lambda$, the Einstein constant  $\kappa$, the Lagrange multiplier $\lambda$, the unit timelike vector field $U^i$,  associated with the velocity four-vector of the aether flow, and the covariant derivative $\nabla_k$ with the connection consistent with the spacetime metric $g_{mn}$, i.e., $\nabla_k g_{mn}=0$. Kinetic terms for the vector and axion fields contain the effective aetheric metric
\begin{equation}
{\cal K}^{ab}_{\ \ mn} = C_1 G^{ab} G_{mn} + C_2 \delta^a_m \delta^b_n + C_3 \delta^a_n \delta^b_m + C_4 U^aU^b G_{mn}  \,,
\label{K}
\end{equation}
\begin{equation}
G^{mn} = g^{mn} + {\cal A} U^m U^n \,,
\label{G}
\end{equation}
where the scalar ${\cal A}(\theta)$ is the guiding function of the first type, and $C_1$, $C_2$, $C_3$, $C_4$ are the Jacobson coupling constants \cite{J1}.
The potential of the axion field $V(\phi, \Phi_*)$ is considered to have the periodic form
\begin{equation}
V(\phi,\Phi_{*}) = \frac{m^2_A \Phi^2_{*}}{2\pi^2} \left[1- \cos{\left(\frac{2 \pi \phi}{\Phi_{*}}\right)} \right] \,,
\label{V}
\end{equation}
where $\Phi_*(\Theta)$ is the guiding function of the second type, and the parameter $\Psi_0$ relates to the coupling constant of the
axion-photon interaction $g_{A \gamma \gamma}$, $\frac{1}{\Psi_0}=g_{A \gamma \gamma}$. The potential (\ref{V}) inherits the discrete symmetry $\frac{2\pi \phi}{\Phi_*} \to \frac{2\pi \phi}{\Phi_*} {+} 2\pi n$. This periodic potential has the minima at $\phi=n \Phi_{*}$. Near the minima, when
$\phi \to n \Phi_{*} {+} \psi$ and $|\frac{2\pi \psi}{\Phi_*}|$ is small, the potential takes the standard form $V \to m^2_A \psi^2$, where $m_A$ is the axion rest mass.
When $\phi {=} n \Phi_*$ ($n$ is an integer), we deal with the axionic analog of the equilibrium state \cite{B2}, since $V_{|\phi{=}n\Phi_*} {=}0$, and $\left(\frac{\partial V}{\partial \phi} \right)_{|\phi{=}n\Phi_*} {=}0$.

The following decompositions are associated with the unit four-vector $U^j$:
\begin{equation}
\nabla_k = U_k D + \nablab_k \,, \quad D = U^s \nabla_s \,, \quad \nablab_k = \Delta_k^j \nabla_j \,, \quad \Delta_k^j = \delta^j_k - U^j U_k \,.
\label{nabla1}
\end{equation}
Here $D$ is the convective derivative, and $\Delta_k^j$ is the projector. The covariant derivative $\nabla_k U_j$ can be decomposed as
\begin{equation}
\nabla_k U_j = U_k DU_j + \sigma_{kj} + \omega_{kj} + \frac13 \Delta_{kj} \Theta \,,
\label{nabla2}
\end{equation}
where the acceleration four-vector $DU_j \equiv a_j$, the symmetric traceless shear tensor $\sigma_{kj}$, the skew - symmetric vorticity tensor $\omega_{kj}$ and the expansion scalar $\Theta$ are presented by the well-known formulas
\begin{equation}
DU_j {=} U^s \nabla_s U_j \,, \quad \sigma_{kj} {=} \frac12 \left(\nablab_k U_j {+} \nablab_j U_k \right) {-} \frac13 \Delta_{kj} \Theta \,, \quad \omega_{kj} {=} \frac12 \left(\nablab_k U_j {-} \nablab_j U_k \right) \,, \quad
\Theta {=} \nabla_kU^k \,.
\label{nabla3}
\end{equation}
The decomposition (\ref{nabla2}) allows us to introduce one linear and three quadratic scalars
\begin{equation}
\Theta = \nabla_k U^k \,, \quad a^2 = DU_k DU^k \,, \quad \sigma^2 = \sigma_{mn} \sigma^{mn} \,, \quad \omega^2 = \omega_{mn} \omega^{mn} \,,
\label{nabla4}
\end{equation}
and thus the kinetic term of the vector field can be rewritten in the form
\begin{equation}
{\cal K}^{ab}_{\ \ mn}(\nabla_a U^m) (\nabla_b U^n) {=}
 [C_1(1{+}{\cal A}) {+} C_4] a^2 {+}
(C_1 {+} C_3)\sigma^2 {+}
(C_1 {-} C_3)\omega^2 {+} \frac13 \left(C_1 {+} 3C_2 {+}C_3 \right) \Theta^2. \label{nabla5}
\end{equation}
Taking into account the constraints obtained after the detection of the event GRB170817 \cite{GRB170817}, we have to put $C_1{+}C_3=0$ into (\ref{nabla5}).

\subsection{Master equations of the model}

\subsubsection{Master equations for the unit vector field}

Variations of the extended action functional (\ref{0}) with respect to the Lagrange multiplier $\lambda$ gives the normalization condition
\begin{equation}
g_{mn}U^m U^n =1 \,.
\label{14}
\end{equation}
Variation with respect to the four-vector $U^i$ gives the aetheric balance equations
\begin{equation}
\nabla_a {\cal J}^{aj}  = \lambda  U^j  {-}  {\cal A}\kappa \Psi^2_0  D \phi \nabla^j \phi  {-} \nabla^j \left(\Omega_1 \frac{d \Phi_*}{d \Theta} {+} \Omega_2 \frac{d {\cal A}}{d \Theta} \right) \,,
\label{15}
\end{equation}
where the following definitions are used:
\begin{equation}
{\cal J}^{aj} = {\cal K}^{abjn} \nabla_b U_n
= C_1 \left( \nabla^a U^j - \nabla^j U^a \right) + C_2 g^{aj} \Theta + \left(C_4+ C_1 {\cal A}\right)U^a DU^j  \,,
\label{16}
\end{equation}
\begin{equation}
\Omega_1 = \frac{\kappa \Psi_0^2 m_A^2}{2\pi^2} \left\{\Phi_* \left[1{-} \cos{\left(\frac{2 \pi \phi}{\Phi_{*}}\right)}\right] {-} \pi \phi \sin{\left(\frac{2 \pi \phi}{\Phi_{*}}\right)} \right\}  \,,
\label{Omega1}
\end{equation}
\begin{equation}
\Omega_2 =  - \frac12 \kappa \Psi^2_0 (D \phi)^2   \,.
\label{Omega2}
\end{equation}
Convolution of (\ref{15}) with $U_j$ gives us the Lagrange multiplier $\lambda$:
\begin{equation}
\lambda =  U_j \nabla_a {\cal J}^{aj}  +  {\cal A}\kappa \Psi^2_0  (D \phi)^2  {+} D \left(\Omega_1 \frac{d \Phi_*}{d \Theta} {+} \Omega_2 \frac{d {\cal A}}{d \Theta} \right)
  \,.
\label{150}
\end{equation}

\subsubsection{Master equation for the axion field}

Variation of the extended action functional (\ref{0}) with respect to the axion field yields
\begin{equation}
\nabla_m \left[\left(g^{mn} + {\cal A} U^m U^n \right)\nabla_n \phi \right] + \frac{m^2_A \Phi_{*}}{2\pi} \sin{\left(\frac{2 \pi \phi}{\Phi_{*}}\right)}  = 0 \,,
\label{24}
\end{equation}
or equivalently
\begin{equation}
 (1{+} {\cal A}) D^2 \phi {+} [(1{+}{\cal A})\Theta + D {\cal A}]D\phi {-} DU^m  \nablab_m \phi {+} \nablab_m \nablab^m \phi {+}
 \frac{m^2_A \Phi_{*}}{2\pi} \sin{\left(\frac{2 \pi \phi}{\Phi_{*}}\right)}  {=} 0 \,.
\label{243}
\end{equation}
Below we use the ansatz that when the axion field is in the equilibrium state, which corresponds to the basic minimum $\phi {=} \Phi_*$, we obtain the master equation for the guiding function of the second type $\Phi_*(\Theta)$, i.e.,
\begin{equation}
\nabla_m \left[\left(g^{mn} + {\cal A} U^m U^n \right)\nabla_n \Phi_* \right] = 0 \,.
\label{247}
\end{equation}

\subsubsection{Master equations for the gravitational field}

Variation of the extended action functional (\ref{0}) with respect to the metric gives the gravity field equations:
\begin{equation}
	R_{ik} - \frac12 R g_{ik} - \Lambda g_{ik} =  T^{(\rm U)}_{ik} + \kappa T^{(\rm A)}_{ik} + T^{(\rm INT)}_{ik} \,.
	\label{25}
\end{equation}
The extended stress-energy tensor of the aether  $T^{(\rm U)}_{ik}$ contains the following elements:
\begin{equation}
	T^{(\rm U)}_{ik} =
	\frac12 g_{ik} \ {\cal K}^{ab}_{\ \ mn} \nabla_a U^m \nabla_b U^n {+} \nabla^m \left[U_{(i}{\cal J}_{k)m} {-}
{\cal J}_{m(i}U_{k)} {-}
{\cal J}_{(ik)} U_m\right]+ U_iU_k U_j \nabla_a {\cal J}^{aj} +
\label{326}
\end{equation}
$$
+C_1\left[(\nabla_mU_i)(\nabla^m U_k) {-}
(\nabla_i U_m )(\nabla_k U^m) \right]
{+}(C_4 + C_1 {\cal A})\left( D U_i D U_k - U_iU_k DU_m DU^m \right)\,.
$$
As usual, the parentheses symbolize the symmetrization of indices.
The extended stress-energy tensor of the axion field is of the form:
\begin{equation}
	T^{(\rm A)}_{ik}= \Psi^2_0 \left[\left(1 {+} {\cal A}\right) {\dot{\phi}}^2 \left( U_iU_k -\frac12 g_{ik}\right)  + \frac12 g_{ik} V  \right]\,.
		\label{T76}
\end{equation}
The part of the total stress-energy tensor associated with the interaction terms contains the derivatives of the guiding functions ${\cal A}$ and $\Phi_*$  with respect to their argument $\Theta$:
\begin{equation}
 T^{(\rm INT)}_{ik} =
-  g_{ik} \Theta  \left(\Omega_1 \frac{d \Phi_*}{d \Theta} {+} \Omega_2 \frac{d {\cal A}}{d \Theta} \right)
{-} \Delta_{ik} \left[ D \left(\Omega_1 \frac{d \Phi_*}{d \Theta} {+} \Omega_2 \frac{d {\cal A}}{d \Theta} \right)\right]
 \,.
\label{T92}
\end{equation}
The Bianchi identity
\begin{equation}
	\nabla^k \left[T^{(\rm U)}_{ik} + \kappa T^{(\rm A)}_{ik}  + T^{(\rm INT)}_{ik} \right] = 0
	\label{T99}
\end{equation}
holds automatically on the solutions to the master equations for the vector and pseudoscalar fields.

\section{Application to the spatially isotropic homogeneous cosmological model}\label{sec3}

\subsection{The spacetime platform, reduced master equations and their solutions}

\subsubsection{Geometric Aspects}

We work below with the spacetime of the Friedmann-Lema\^itre-Robinson-Walker type, with the metric
\begin{equation}
	ds^2 = dt^2 - a^2(t)\left(dx^2 + dy^2 + dz^2 \right) \,.
	\label{m1}
\end{equation}
The velocity four-vector of the aether flow is known to be of the form $U^j = \delta^j_0$, and the corresponding covariant derivative of the vector field has the following decomposition
\begin{equation}
	\nabla_k U_i = \frac12 \dot{g}_{ik} = \frac{\dot{a}}{a} \Delta_{ik} = H \Delta_{ik} = \frac13 \Theta \Delta_{ik} \,.
	\label{m2}
\end{equation}
Clearly, in this case  $DU_j=0$, $\sigma_{mn}=0$, $\omega_{mn}=0$, $\Theta = 3 H = 3 \frac{\dot{a}}{a}$, and standardly
the dot denotes the derivative with respect to the cosmological time $t$.

\subsubsection{Solution to the equations of the vector field}

Keeping in mind that $DU_j{=}0$, $\sigma_{mn}{=}0$, $\omega_{mn}{=}0$ we find that the extended Jacobson's tensor (\ref{16}) converts into
\begin{equation}
	J^{aj} = C_2 \Theta  g^{aj}  \,,
	\label{JFRiedmann}
\end{equation}
and the equations for the unit vector field (\ref{15}) takes the form
\begin{equation}
	C_2 \nabla_j \Theta  = \lambda  U_j  - \kappa \Psi^2_0 {\cal A}U_j {\dot{\phi}}^2 -\nabla_j \left(\Omega_1 \frac{d \Phi_*}{d \Theta} + \Omega_2 \frac{d {\cal A}}{d \Theta}\right) \,.
	\label{N15}
\end{equation}
Four equations (\ref{N15}) contain only one non-trivial equation, which gives the solution for the Lagrange multiplier $\lambda$:
\begin{equation}
\lambda = 	C_2 \dot{\Theta} + \kappa \Psi^2_0 {\cal A} {\dot{\phi}}^2 + \frac{d}{dt} \left(\Omega_1 \frac{d \Phi_*}{d \Theta} + \Omega_2 \frac{d {\cal A}}{d \Theta}\right) \,.
	\label{N15v}
\end{equation}
Thus, the aetheric subset of the total system of master equations is solved.

\subsubsection{First integral of the reduced equation for the axion field}

We suppose that the axion field $\phi$ is frozen in the first minimum of the axion potential, i.e., $\phi {=} \Phi_*(t)$. Then we put $\phi {=} \Phi_*$ into (\ref{243}) and obtain the key equation for $\Phi_*(t)$
\begin{equation}
 (1{+} {\cal A}) \ddot{\Phi}_* {+} \left[3(1{+}{\cal A})\frac{\dot{a}}{a} + \dot {\cal A}\right] \dot{\Phi}_* = 0 \,,
\label{243v}
\end{equation}
which admits the first integral
\begin{equation}
\dot{\Phi}_*(t) = \frac{\rm const}{a^3(t) [1+{\cal A}(t)]} = \dot{\Phi}_*(t_0) \left[\frac{a(t_0)}{a(t)}\right]^3 \frac{\left[1+{\cal A}(t_0)\right]}{\left[1+{\cal A}(t)\right]} \,.
\label{keyPhi}
\end{equation}
The parameter $t_0$ describes the initial time moment; ${\cal A}(t_0)$ is the initial value of the guiding function of the first type, and $\dot{\Phi}_*(t_0)$ indicates the initial value of the first derivative of the guiding function of the second type.

\subsubsection{Key equations for the gravity field }

When $\phi {=} \Phi_*$, the function $\Omega_1$ takes zero value, and the reduced extended equations of the gravitational field converts into one key equation
\begin{equation}
	\frac13 \Theta^2\left(1 + \frac32 C_2 \right) - \Lambda =    \frac12 \kappa \Psi^2_0 {\dot{\Phi}_*}^2 \left[1+ {\cal A} + \Theta \frac{d {\cal A}}{d\Theta} \right]
	\,.
	\label{0G1}
\end{equation}
Since $\dot{\Phi}_*$ is already found and is of the form (\ref{keyPhi}), we obtain the equation, which connects the scalar $\Theta$ with the reduced scale factor $x= \frac{a(t)}{a(t_0)}$ as follows:
\begin{equation}
	\frac13 \Theta^2\left(1 + \frac32 C_2 \right) - \Lambda =    \frac{1}{2x^6} \kappa \Psi^2_0 {\dot{\Phi}_*}^2(t_0) \left[1+ {\cal A}(t_0)\right]^2 \left[\frac{1}{1+ {\cal A}} - \Theta \frac{d }{d \Theta}
\left(\frac{1}{1+ {\cal A}}\right) \right]
	\,.
	\label{G1}
\end{equation}
Then we assume that $C_2> -\frac23$, $\Lambda >0$, and introduce the auxiliary parameters
\begin{equation}
 H_{\infty} = \sqrt{\frac{\Lambda}{3(\left(1 + \frac32 C_2 \right))}}
\,, \quad
h^2 = \frac{\kappa \Psi^2_0 {\dot{\Phi}}^2(t_0) \left[1+ {\cal A}(t_0)\right]^2}{6 \left(1 + \frac32 C_2 \right)} \,.
\label{OKH}
\end{equation}
Now we are ready to analyze the main equation of the model for the function $H(x)$
\begin{equation}
x^6 \left[H^2 - H^2_{\infty}\right] =   h^2 \left[\frac{1}{1+ {\cal A}} - H \frac{d }{d H}
\left(\frac{1}{1+ {\cal A}}\right) \right] \,.
\label{OKH2}
\end{equation}

\subsection{Modeling of the guiding function of the first type}

When we discuss the structure of the guiding function of the first type we use two assumptions. First, we assume that ${\cal A} {=} 0$, if $\Theta=0$. Second, we assume that the right-hand side of the equation (\ref{OKH2}) is a regular function of its argument $H$, and thus we can use the decomposition
\begin{equation}
\left[\frac{1}{1+ {\cal A}} - H \frac{d }{d H}
\left(\frac{1}{1+ {\cal A}}\right) \right] =   1 -  \gamma_1 H -  \gamma_2 H^2 - 2 \gamma_3 H^3 - 3\gamma_4 H^4 -...
\label{G01}
\end{equation}
This decomposition allows us to reconstruct the function $\frac{1}{1{+}{\cal A}}$, which has the form
\begin{equation}
\frac{1}{1+{\cal A}} = 1 +  \gamma_1 H \left[1+ \log{\frac{H}{H_*}} \right] + \gamma_2 H^2 + \gamma_3 H^3 + \gamma_4 H^4 + ...
\label{G2}
\end{equation}
Here $H_*$ is some constant of integration.
The key to our consideration is the analysis of the asymptotic regime ($x \to \infty$ ) of the equation
\begin{equation}
	x^6 \left[H^2 - H^2_{\infty}\right] =  h^2 \left[1 - \gamma_1 H - \gamma_2 H^2 - 2 \gamma_3 H^3 - 3 \gamma_4 H^4  - ...\right]
	\,.
	\label{G10}
\end{equation}
If we restrict our-selves by the term $H^m$ in the right-hand side of (\ref{G10}), we see that, first,
$H^{m-2} \propto x^6$, second, $H \propto x^{\frac{6}{m-2}}$ and third, $a(t) \propto t^{-\frac{m-2}{6}}$. In other words, if $m>2$, the Universe collapses asymptotically, and this detail is in contradiction with the main idea of perpetual expansion. Of course, this point is disputable, but we follow this idea. Now we deal with the quadratic equation with respect to $H$
\begin{equation}
	x^6 \left[H^2 - H^2_{\infty}\right] =   h^2 \left[1 - \gamma_1 H - \gamma_2 H^2 \right]
	\,,
	\label{0G21}
\end{equation}
and its positive solution is
\begin{equation}
H(x) = \sqrt{\frac{\gamma_1^2 h^4}{4(x^6+ \gamma_2 h^2)^2} +
\frac{H^2_{\infty} x^6 + h^2}{x^6 + \gamma_2 h^2}} - \frac{\gamma_1 h^2}{2(x^6+ \gamma_2 h^2)}
	\,.
	\label{G109}
\end{equation}
With this function $H(x)$ one can reconstruct the scale factor as the function of time, if we use the formal quadrature
\begin{equation}
t-t_0 = \int_1^{\frac{a(t)}{a(t_0)}} \frac{dx}{x H(x)} \,.
\label{G91}
\end{equation}
Clearly, there are two asymptotic regimes.

1) When $\Lambda \neq 0$, $H \to H_{\infty}$ and thus $a(t) \propto e^{H_{\infty}t}$.

2) When $\Lambda {=} 0$, $H \propto \frac{1}{x^3}$ and thus $a(t) \propto t^{\frac13}$.

In order to have further progress in calculations, we consider three analytically solvable submodels.

\subsubsection{First analytically solvable submodel}

Let us consider the model with $\gamma_1 = - \frac{1}{H_{\infty}}$ and $\gamma_2=0$.
In this case the function ${\cal A}(H)$ satisfies the relationship
\begin{equation}
\frac{1}{1+{\cal A}} = 1 - \frac{H}{H_{\infty}} \left[1+ \log{\frac{H}{H_*}} \right] \,.
\label{G31}
\end{equation}
In order to simplify the analysis, we assume that $H_*= H_{\infty}$ and obtain the following expression for  the guiding function of the first type
\begin{equation}
{\cal A} = \frac{\frac{H}{H_{\infty}} \left(1 +\log{\frac{H}{H_{\infty}}} \right)}
{1 - \frac{H}{H_{\infty}} \left(1 + \log{\frac{H}{H_{\infty}}}\right)} \,.
\label{G32}
\end{equation}
Formally speaking, this function takes the infinite value, when the denominator is equal to zero. But this situation appears only at infinity $a=\infty$, when $H=H_{\infty}$.
Now we deal with the key equation
\begin{equation}
H^2 - H^2_{\infty} =   \frac{h^2}{H_{\infty} x^6} \left(H_{\infty} +  H \right)
	\,,
	\label{G21}
\end{equation}
we omit the negative root $H=-H_{\infty}$, and see that the positive solution is
\begin{equation}
H(x) = H_{\infty} + \frac{h^2}{H_{\infty} x^6} \,.
\label{G41}
\end{equation}
Mention should be made, that this model is self-consistent, when, first,  $H(t_0)>H_{\infty}$, second, $h^2{=} H_{\infty}[H(t_0){-}H_{\infty}]$. According to the definition (\ref{OKH}) the last requirement links the values ${\cal A}(t_0)$, $\dot{\Phi}(t_0)$ and $H(t_0)$.

The scale factor $a(t)$ and the Hubble function $H(t)$  can be now presented in the form
\begin{equation}
a(t) = a(t_0) \left[\left(1+ \frac{h^2}{H^2_{\infty}} \right) e^{6H_{\infty}(t-t_0)} - \frac{h^2}{H^2_{\infty}} \right]^{\frac16} \,,
\label{G51}
\end{equation}
\begin{equation}
H(t) = \frac{H_{\infty}}{\left\{1-\left[1-\frac{H_{\infty}}{H(t_0)} \right] e^{-6H_{\infty}(t-t_0)} \right\}} \,.
\label{G521}
\end{equation}
The acceleration parameter $-q(t)$ given by the formula
\begin{equation}
-q(t) =  \frac{\ddot{a}}{a H^2} =  1- \left(\frac{6 h^2}{h^2 + H^2_{\infty}}\right) e^{-6H_{\infty}(t-t_0)}
\label{G199}
\end{equation}
is the monotonic function of time, and it tends to one asymptotically at $t \to \infty$.

Finally, we intend to reconstruct the guiding function of the second type $\Phi_*(H)$. The simplest way is the following. First, using the replacements $t \to x =\frac{a(t)}{a(t_0)}$ and $ \frac{d}{dt} \to  x H(x)\frac{d}{dx}$,  we rewrite the relationship (\ref{keyPhi}) as follows
\begin{equation}
\Phi^{\prime}_*(x) = - \frac{\dot{\Phi}_*(t_0) \left[1+{\cal A}(t_0)\right]}{H_{\infty}x^4} \left[- \frac{H_{\infty}}{H} + 1+ \log{\left(\frac{H}{H_{\infty}}\right)}  \right] \,.
\label{3N7}
\end{equation}
Second, using (\ref{G41}), we integrate (\ref{3N7}) and obtain
\begin{equation}
\Phi_*(x) = \Phi_*(t_0) {+} \frac{\dot{\Phi}_*(t_0) \left[1{+}{\cal A}(t_0)\right]}{3H_{\infty}} \  \Re_1(x) \,,
\label{3N8}
\end{equation}
\begin{equation}
\Re_1(x) \equiv
 \left(1{-}\frac{1}{x^3}\right) {+}
\frac{1}{x^3}\log{\left(1{+} \frac{h^2}{H^2_{\infty} x^6} \right)} {-}   \log{\left(1+ \frac{h^2}{H^2_{\infty}} \right)} {+}
\label{9N8}
\end{equation}
$$
+ \frac{H_{\infty}}{|h|} \left(\arctan{\frac{|h|}{H_{\infty}x^3}} {-} \arctan{\frac{|h|}{H_{\infty}}} \right) \,.
$$
Third, using the replacement $\frac{1}{x^6} = \frac{H_{\infty}}{h^2}(H-H_{\infty})$, we recover the function $\Phi_*(H)$ based on the solution (\ref{3N8}). Asymptotic value of the reconstructed guiding function is
\begin{equation}
\Phi_*(\infty) = \Phi_*(t_0) + \frac{\dot{\Phi}_*(t_0) \left[1+{\cal A}(t_0)\right]}{3H_{\infty}} \Re_1(\infty)\,,
\label{3N9}
\end{equation}
$$
\Re_1(\infty) =	 1 {-} \log{\left(1+ \frac{h^2}{H^2_{\infty}} \right)}  {-} \frac{H_{\infty}}{|h|} \arctan{\frac{|h|}{H_{\infty}}}    \,.
$$
Figure 1 illustrates the details of the function $\Re_1(x)$.
\begin{figure}[ht]
	\includegraphics[width=10.5 cm, height = 7cm]{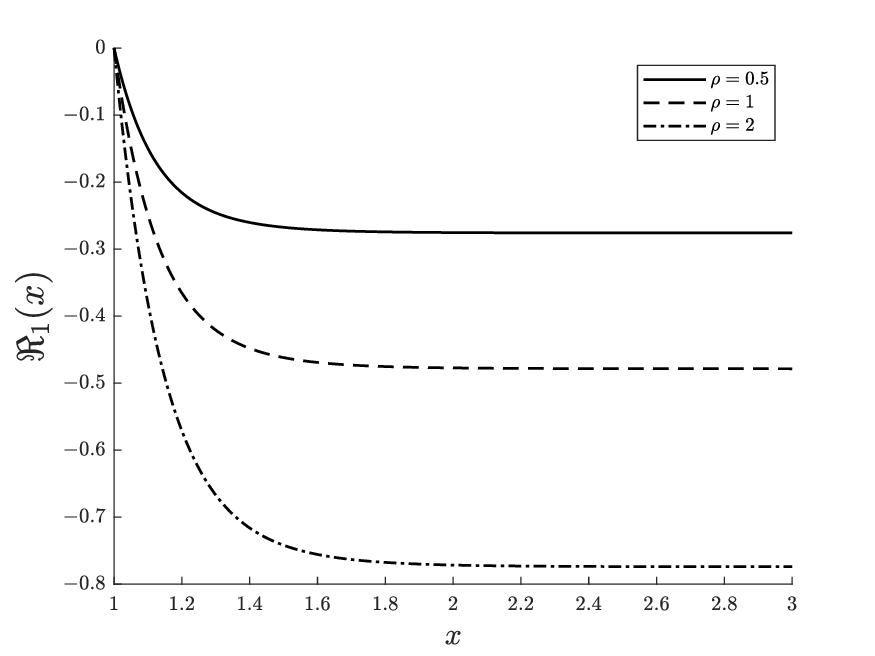}
	\caption{Illustration of the behavior of the function $\Re_1(x)$ (\ref{9N8}), which enters the guiding function of the second type $\Phi_*$, for three values of the parameter
$\rho {=} \frac{|h|}{H_{\infty}}$. All the curves start with the value $\Re(1)=0$ and tend monotonically to their asymptotic values $\Re_1(\infty)$ (\ref{3N9}). \label{fig1}}
\end{figure}

\subsubsection{Second analytically solvable submodel}

The second submodel relates to the case, when  $\Lambda \neq 0$, $\gamma_1=0$ and $\gamma_2=\frac{\alpha^2}{H^2_{\infty}} >0$.
With these assumptions the guiding function of the first type
\begin{equation}
{\cal A}(H) = - \frac{\gamma_2 H^2}{1+ \gamma_2 H^2} = - \frac{\alpha^2 H^2}{H^2_{\infty}+ \alpha^2 H^2}
\label{M21}
\end{equation}
is the regular function of the Hubble function $H$. From the key equation for the gravity field (\ref{0G21}) we obtain
\begin{equation}
H(x) = H_{\infty} \sqrt{\frac{x^6 + \frac{h^2}{H^2_{\infty}}}{x^6 + \frac{\alpha^2 h^2}{H^2_{\infty}}}} \,.
\label{M22}
\end{equation}
The parameter $\alpha^2$ is connected with the initial value of the Hubble function as follows:
\begin{equation}
H(t_0) \equiv H(x=1) = H_{\infty} \sqrt{\frac{1 + \frac{h^2}{H^2_{\infty}}}{1 + \frac{\alpha^2 h^2}{H^2_{\infty}}}} \,.
\label{M23}
\end{equation}
Clearly, we have to distinguish the cases $\alpha^2=1$ and $\alpha^2 \neq 1$.

1) When $\alpha^2=1$, we obtain that the Hubble function converts into the constant $H(x)=H(1)=H_{\infty}$, and we deal with
the de Sitter type behavior of the Universe, for which $a(t)=a(t_0)e^{H_{\infty}(t-t_0)}$.
The guiding function of the first type also is constant ${\cal A} = -\frac12$, and the guiding function of the second type behaves as
\begin{equation}
\Phi_*(t) = \Phi_*(t_0) - \frac{{\dot{\Phi}_*}^2(t_0) a^3(t_0)}{3H_{\infty}} e^{-3H_{\infty}(t-t_0)} \,.
\label{M25}
\end{equation}

2) When $\alpha^2 \neq 1$, direct integration of (\ref{G91}) yields
\begin{equation}
e^{6H_{\infty}(t-t_*)} = \left|\frac{(z-\alpha)^{\alpha}(z+1)}{(z+\alpha)^{\alpha}(z-1)} \right| \,,
\label{M26}
\end{equation}
where we used the positive root $\alpha = + \sqrt{\alpha^2}$. The auxiliary function $z(t)$ and two new parameters, $z_*$ and $t_*$ are:
\begin{equation}
z = \sqrt{\frac{H^2_{\infty}\left[\frac{a(t)}{a(t_0)}\right]^6+ \alpha^2 h^2}{H^2_{\infty}\left[ \frac{a(t)}{a(t_0)}\right]^6+ h^2 }} \,, \quad z_* = \sqrt{\frac{H^2_{\infty}+ \alpha^2 h^2}{H^2_{\infty}+h^2 }} \,,
\label{M27}
\end{equation}
\begin{equation}
t_* = t_0 - \frac{1}{6H_{\infty}}\log{\left[\frac{(z_*+1)(z_*-\alpha)^{\alpha}}{(z_*-1)(z_*+\alpha)^{\alpha}} \right]} \,.
\label{M28}
\end{equation}
According to (\ref{M27}) $z \to 1$, when $a \to \infty$; the corresponding asymptotic behavior is characterized by the de Sitter type law
\begin{equation}
a(t, \alpha)  \to a(t_0) \left(\frac{h}{2H_{\infty}}\right)^{\frac13} \left|\frac{1+\alpha}{1-\alpha} \right|^{\frac{\alpha-1}{6}} e^{H_{\infty}(t-t_*)}  \,.
\label{M287}
\end{equation}
The formulas (\ref{M26}), (\ref{M27}) and (\ref{M28}) give us the implicit representation. The function $a(t)$ has no extrema; we have illustrated the behavior of the scale factor in the early epoch on Figure 2.

\begin{figure}[ht]
	\includegraphics[width=10.5 cm, height = 7cm]{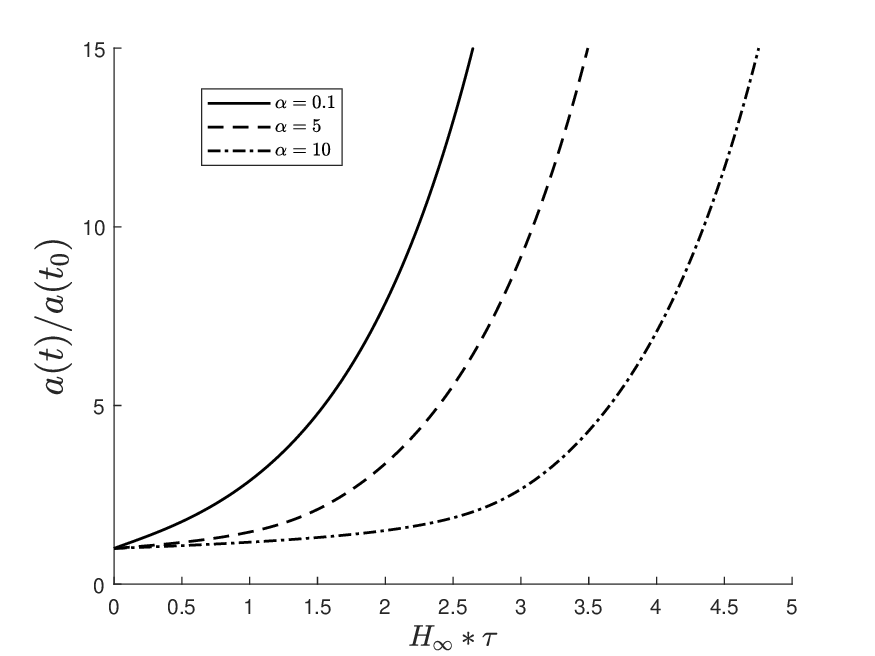}
	\caption{Illustration of the behavior of the reduced scale factor $\frac{a(t)}{a(t_0)}$ in the early epoch; this function is presented in the implicit form by (\ref{M26}). Here $\tau {=} t{-}t_0$. \label{fig3}}
\end{figure}

 The guiding function of the second type can be represented in terms of elliptic functions. For instance, if $0 < \alpha <1$ the term
\begin{equation}
\Phi_*(x)= \Phi_*(t_0) - \frac{\dot{\Phi}_*(t_0) \left[1+{\cal A}(t_0)\right]}{3H_{\infty}} \ \Re_2(x)
\label{M29}
\end{equation}
contains the special function $\Re_2(x)$, which is equal to
\begin{equation}
\Re_2(x) = \int_1^{\frac{1}{x^3}} dz \left[\sqrt{\frac{1 + \alpha^2 \frac{h^2}{H^2_{\infty}} z^2}{1 + \frac{h^2}{H^2_{\infty}}z^2}}  +
\alpha^2 \sqrt{\frac{1 + \frac{h^2}{H^2_{\infty}}z^2}{1 + \alpha^2 \frac{h^2}{H^2_{\infty}} z^2} } \right] =
\label{M30}
\end{equation}
$$
= \frac{H_{\infty}}{h} \left\{(1+\alpha^2)\left[F(\varphi,k) {-} F\left(\varphi_*,k \right) \right] {-} 2\left[E(\varphi,k) {-} E\left(\varphi_*,k \right) \right]\right\} {+}
$$
$$
{+} \frac{2}{x^3} \sqrt{\frac{H^2_{\infty} x^6 {+} \alpha^2 h^2}{H^2_{\infty} x^6 {+} h^2}} {-} 2\sqrt{\frac{H^2_{\infty}{+} \alpha^2 h^2}{H^2_{\infty} {+} h^2}} \,,
$$
where the elliptic functions of the first and second types, respectively,
\begin{equation}
F(\varphi,k) \equiv \int_0^{\varphi} \frac{d \psi}{\sqrt{1-k^2 \sin^2{\psi}}} \,, \quad
E(\varphi,k) \equiv \int_0^{\varphi} d \psi \sqrt{1-k^2 \sin^2{\psi}}
\label{M32}
\end{equation}
are characterized by the arguments
\begin{equation}
\varphi = \arctan{\left(\frac{h}{H_{\infty} x^3}\right)} \,, \quad \varphi_* = \arctan{\left(\frac{h}{H_{\infty}}\right)} \,, \quad k = \sqrt{1-\alpha^2} \,.
\label{M33}
\end{equation}
The asymptotic value of the guiding function of the second type is
\begin{equation}
\Phi_*(x) {=} \Phi_*(t_0) {+} \frac{\dot{\Phi}_*(t_0) \left[1{+}{\cal A}(t_0)\right]}{3H_{\infty}} \left\{ \frac{H_{\infty}}{h} \left[(1{+}\alpha^2) F\left(\varphi_*,k \right)
 {-} 2E \left(\varphi_*,k \right) \right] {+} 2\sqrt{\frac{H^2_{\infty}{+} \alpha^2 h^2}{H^2_{\infty} {+} h^2}}    \right\} \,.
\end{equation}

\subsubsection{Third analytically solvable submodel}

Now we assume that the cosmological constant is equal to zero, $\Lambda {=} 0$, i.e., $H_{\infty}{=}0$. Also we assume that $\gamma_1=0$ and $\gamma_2=\frac{\nu^6}{h^2}>0$.
We obtain again that ${\cal A}(H)$ is regular
\begin{equation}
{\cal A}(H) = - \frac{\nu^6 H^2}{h^2+ \nu^6 H^2} \,,
\label{M31}
\end{equation}
and the Hubble function is of the form
\begin{equation}
H(x) = \frac{|h|}{\sqrt{x^6 + \nu^6}}	\,.
	\label{M32}
\end{equation}
Then we obtain the reduced scale factor $x(t)$ in the implicit form
\begin{equation}
\frac{3|h|}{\nu^3}(t-t_{**}) = \sqrt{1 + \frac{x^6}{\nu^6}} - \log{\left[\sqrt{1+\frac{\nu^6}{x^6}}+ \frac{\nu^3}{x^3}\right]} \,,
	\label{M33}
\end{equation}
where we introduced for simplicity the formal parameter $t_{**}$
\begin{equation}
t_{**} = t_0 - \frac{1}{3|h|}\sqrt{1+ \nu^6} - \frac{\nu^3}{3|h|} \log{\left(\sqrt{1 + \nu^6}-\nu^3 \right)} \,.
	\label{M34}
\end{equation}
Finally, we obtain the guiding function of the second type as the function of the reduced scale factor
\begin{equation}
\Phi_*(x) = \Phi_*(t_0) + \frac{1}{3 |h|} \dot{\Phi}_*(t_0) \left[1+{\cal A}(t_0)\right] \ \Re_3(x) \,.
\label{M39}
\end{equation}
$$
 \Re_3(x) \equiv \log{\left[\frac{\left(x^3 {+} \sqrt{\nu^6 {+} x^6}\right)}{\left(1 {+} \sqrt{1{+} \nu^6}\right)}\right]} - 2\sqrt{1+ \frac{\nu^6}{x^6}} + 2\sqrt{1+ \nu^6}
 \,.
$$
In the asymptotic limit $x \to \infty$ the function $\Phi_*(H)$ has the form
\begin{equation}
\Phi_*(H) = \Phi_*(t_0) - \frac{1}{3 |h|} \dot{\Phi}_*(t_0) \left[1+{\cal A}(t_0)\right] \log{\left(\frac{\nu^3 H}{2|h|}\right)}
 \,.
\label{M310}
\end{equation}
Figure 3 illustrates the behavior of the function $\Re_{3}(x)$.

\begin{figure}[ht]
	\includegraphics[width=10.5 cm, height = 7cm]{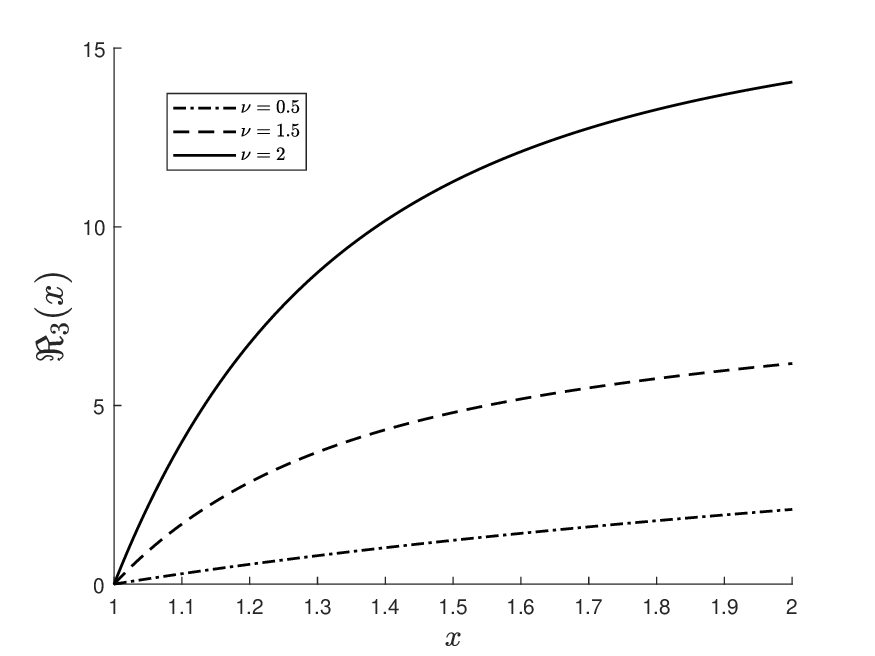}
	\caption{Illustration of the behavior of the function $\Re_3(x)$ for three values of the parameter $\nu$.\label{fig5}}
\end{figure}

\subsubsection{Special case}

The last interesting submodel relates to the case ${\cal A}= {-}1$, for which the aetheric effective metric converts into the projector $G^{mn} \to \Delta^{mn}=g^{mn} {-} U^m U^n$. For such guiding function of the first type the axion field equation (\ref{243}) admits the solution depending on time, if and only if $\phi {=} n \Phi_* $, and thus $V{=}0$. The equation for the gravity field (\ref{0G1}) gives the de Sitter type solution $H{=}H_{\infty}$, and the equation (\ref{243v}) turns into the identity $0{=}0$. In other words, the guiding function of the second type happens to be arbitrary constant $\Phi_*(t){=} \Phi_*(H_{\infty})$.

\section{Discussion and conclusions}\label{sec4}

In the presented work we studied new exact solutions to the master equations of the extended version of the Einstein-aether-axion theory. The main idea of the theory extension is based on the introduction of two guiding functions ${\cal A}(\Theta)$ and $\Phi_{*}(\Theta)$, which depend on the expansion scalar of the aether flow, $\Theta {=} \nabla_k U^k$. This choice is dictated by the fact that within the Friedmann-Lema\^itre-Robinson-Walker model there is only one non-vanishing invariant reconstructed using the covariant derivative $\nabla_k U^j$ of the aether velocity four-vector $U^j$. The bonus of this approach is that in the FLRW model $\Theta = 3H$, and thus the aetheric control over the axion system evolution happens to be described in terms of the Hubble function $H(t)$, which is intrinsic for this model and has clear physical meaning. Why we used namely two guiding functions? We kept in mind that generally the axion system is characterized by two state functions: kinetic and potential energy. Modification of the kinetic term in the Lagrangian of the extended theory is performed using the effective aetheric metric $G^{mn} {=} g^{mn} {+} {\cal A} U^m U^n$ (see (\ref{0})), where the scalar ${\cal A}(\Theta)$ has been indicated as the guiding function of the first type. Modification of the  axion field potential is made by the introduction of the guiding function of the second type $\Phi_*(\Theta)$, which predetermines the location and depth of the potential minima (see (\ref{V})).

The next question is how one can find ${\cal A}(\Theta)$ and $\Phi_*(\Theta)$? We have proposed the following idea. If the axion field is frozen in the first minimum of the potential, i.e., is in the first equilibrium state $\phi {=} \Phi_* $, we see that the corresponding equation of the axion field (see (\ref{247}) and (\ref{243v})) can be indicated as the master equation for the guiding function of the second type. Fortunately, the equation  (\ref{243v})) admits the first integral (\ref{keyPhi}), which can be put into the equations for the gravity field, thus providing the key equation (\ref{0G1}) to be self-closed equation for the scalar function $\Theta(x)$, or equivalently, for the Hubble function $H(x)$. When $H$ is found, the guiding function of the second type $\Phi_*$ can be reconstructed by the direct integration (see the results (\ref{3N8}), (\ref{M29}), (\ref{M30}) and (\ref{M39})).

Regarding the search for the guiding function of the first type ${\cal A}(\Theta)$, we follow the idea that, first, the right-hand side of the key equation of the gravity field (\ref{OKH2}) has to be a regular function, second, the model has to describe the perpetual Universe expansion without Big Rip and Big Crunch. From these two requirements we restore the function ${\cal A}(H)$ up to three arbitrary parameters $\gamma_1$, $\gamma_2$ and $H_*$ using the formula
$$
\frac{1}{1+{\cal A}} = 1 +  \gamma_1 H \left[1+ \log{\frac{H}{H_*}} \right] + \gamma_2 H^2 \,.
$$
The Hubble function $H(x)$ is the solution to the quadratic equation and its positive root has the form (\ref{G109}) for arbitrary parameters $\gamma_1$, $\gamma_2$ and $H_*$; only the scale factor as the function of cosmological time $a(t)$ can be presented in quadratures. In order to obtain results presented in the analytical and special functions, we considered four particular submodels, selecting the listed parameters in a specific way. Research objectives achieved.

The last point of discussion is connected with an application of the extended model for the interpretation of observational data, in particular, for the estimation of the axion mass. In this context, we would like to attract attention to the equation of the axion field evolution (\ref{243}). When the value of the axion field is close to one of the potential minima, i.e., $\phi \to n \Phi_* {+} \psi $ with $\left|\frac{2 \pi \psi}{\Phi_*}\right| <<1$, we deal with the linear differential equation, in which the quantity $M(\Theta) {=} \frac{m_{\rm A}}{\sqrt{1{+}{\cal A}}}$ plays the role of an effective axion mass depending on the scalar of expansion of the aether flow  $\Theta$. Preliminary analysis shows that for some choice of the guiding function ${\cal A}(\Theta)$ this equation admits instable solutions, which are associated with the axionization of the early Universe in analogy with the results obtained in \cite{B3}. The growth of the number of axions in the early Universe leads to the formation of the axionic dark matter detected in our epoch; thus, the parameters of the presented extended model could be linked with the mass density of the relic axions.   Clearly, this part of work should be detailed; it is beyond the scope of this article and is planned to form material for the next publication.

\acknowledgments{This research was funded by Russian Science Foundation.}


\section*{References}

\begin{thebibliography}{99}


\bibitem{GW2015}  Abbott, B.P.; et al. (LIGO Scientific Collaboration and Virgo Collaboration), Observation of Gravitational Waves from a Binary Black Hole Merger. {\em Phys. Rev. Lett. } {\bf 2016}, {\em 116}, 061102.

\bibitem{GW2019} Abbott, R.; et. al.  GW190521: A Binary Black Hole Merger with a Total Mass of 150 M(Sun).  {\em Phys. Rev. Lett.} {\bf 2020}, {\em 125}, 101102.

\bibitem{2022} Pechetti, R.; et.al. Detection of a 100,000 M(Sun) black hole in M31's Most Massive Globular Cluster: A Tidally Stripped Nucleus.
{\em  Ap. J} {\bf 2022}, {\em  924}, 48.

\bibitem{JWST1} Diego, J.M.; et. al. JWST's PEARLS: Mothra, a new kaiju star at z=2.091 extremely magnified by MACS0416, and implications for dark matter models.
{\em Astronomy and Astrophysics} {\bf 2023}, {\em  679}, A31.

\bibitem{JWST2} Spilker, J.S.; et.al. Spatial variations in aromatic hydrocarbon emission in a dust-rich galaxy. {\em Nature} {\bf 2023}, {\em 618}, 708-711.

\bibitem{JWST3} https://www.jameswebbdiscovery.com/


\bibitem[Calderon(2022)]{1} Calderon, R.; L'Huillier, B.; Polarski, D.; Shafieloo, A.; Starobinsky, A.A.
Joint reconstructions of growth and expansion histories from stage-IV surveys with minimal assumptions: Dark Energy beyond?
{\em Phys. Rev. D} {\bf 2022}, {\em 106},  083513.

\bibitem[Calderon(2023)]{2} Calderon, R.; L'Huillier, B.; Polarski, D.; Shafieloo, A.; Starobinsky, A.A.
Joint reconstructions of growth and expansion histories from stage-IV surveys with minimal assumptions. II. Modified gravity and massive neutrinos.
{\em Phys. Rev. D} {\bf 2023}, {\em 108}, 023504.

\bibitem[Capozziello(2006)]{3} Capozziello, S.; Nojiri, S.; Odintsov, S.D.
Unified phantom cosmology: inflation, dark energy and dark matter under the same standard. {\em Phys. Lett. B} {2006}, {\em 632}, P. 597--604.

\bibitem[Sotiriou(2006)]{4} Sotiriou, T.P. Unification of inflation and cosmic acceleration in the Palatini formalism.  {\em Phys. Rev. D} {\bf 2006}, {\em 73}, 063515.

\bibitem[Nojiri(2020)]{5} Nojiri,S.; Odintsov, S.D.; Oikonomou,  V.K. Unifying inflation with early and late-time dark energy in $F(R)$ gravity.
{\em Physics of the Dark Universe} {\bf 2020}, {\em 29}, 100602.

\bibitem[Odintsov(2019)]{6} Odintsov, S.D,; Oikonomou, V.K. Unification of inflation with dark energy in $f(R)$ Gravity and axion dark matter.
{\em Phys. Rev. D} {\bf 2019}, {\em 99}, 104070.

\bibitem[Oikonomou(2021)]{7} Oikonomou, V.K. Unifying of inflation with early and late dark energy epochs in axion $F(R)$ gravity. {\em Phys. Rev. D} {\bf 2021}, {\em 103}, 044036.

\bibitem[Jacobson(2001)]{J1}
Jacobson, T.; Mattingly, D. Gravity with a dynamical preferred frame. {\em Phys. Rev. D} {\bf 2001}, {\em 64}, 024028.

\bibitem[Jacobson(2007)]{J2}
Jacobson, T. Einstein-aether gravity: a status report. {\em PoSQG-Ph} {\bf 2007}, {\em 020}, 020.

\bibitem[Jacobson(2004)]{J3}
Jacobson, T.; Mattingly, D. Einstein-aether waves. {\em Phys. Rev. D} {\bf 2004}, {\em 70}, 024003.

\bibitem[Heinicke(2005)]{J4}
Heinicke, C.; Baekler, P.; Hehl, F.W. Einstein-aether theory, violation of Lorentz invariance, and metric-affine gravity. {\em Phys. Rev. D} {\bf 2005}, {\em 72}, 025012.

\bibitem[Balakin(2019)]{B1}
Balakin, A.B.; Shakirzyanov, A.F. Axionic extension of the Einstein-aether theory: How does dynamic aether regulate the state of axionic dark matter? {\em Physics of the Dark Universe} {\bf 2019}, {\em 24}, 100283.

\bibitem[Balakin(2020)]{B2}
Balakin, A.B.; Shakirzyanov, A.F. Is the axionic Dark Matter an equilibrium System? {\em Universe} {\bf 2020}, {\em 6 (11)}, 192.

\bibitem[Balakin(2023)]{B3} Balakin, A.B.; Ilin, A.S.; Shakirzyanov, A.F. Interaction of the Cosmic Dark Fluid with Dynamic Aether: Parametric Mechanism of Axion Generation in the Early Universe.
{\em Symmetry} {\bf 2023}, {\em 15}, 1824.

\bibitem[Balakin(2021)]{B4} Balakin, A.B.; Efremova, A.O. Interaction of the axionic dark matter, dynamic aether, spinor and gravity fields as an origin of oscillations of the fermion effective mass.
{\em Eur. Phys. J C} {\bf 2021}, {\em 81}, 674.

\bibitem[Balakin(2023)]{B5}Balakin, A.B.; Efremova, A.O. Dynamic aether as a trigger for spontaneous spinorization in early Universe. {\em Universe} {\bf 2023}, {\em 9}, 481.

\bibitem[Balakin(2023)]{Gcos} Balakin, A.B.; Shakirzyanov, A.F. The extended Einstein-Maxwell-aether-axion theory: Effective metric as an instrument of the aetheric control over the axion dynamics.
{\em Gravitation and Cosmology} {\bf 2024}, {\em 30}, No. 1, pp. 57-67.

\bibitem[Peccei(1977)]{Ax1}
Peccei, R.D.; Quinn, H.R. CP conservation in the presence of instantons. {\em Phys. Rev. Lett.} {\bf 1977}, {\em 38}, 1440--1443.

\bibitem[Weinberg(1978)]{A2}
Weinberg, S. A new light boson? {\em Phys. Rev. Lett.} {\bf 1978}, {\em 40}, 223--226.

\bibitem[Wilczek(1978)]{A3}
Wilczek, F. Problem of strong P and T invariance in the presence of instantons. {\em Phys. Rev. Lett.} {\bf 1978}, {\em 40}, 279--282.

\bibitem[Wei-Tou(1977)]{A4}
Ni, Wei-Tou. Equivalence principles and electromagnetism. {\em Phys. Rev. Lett.} {\bf 1977}, {\em 38}, 301--304.

\bibitem[Sikivie(1983)]{A5}
Sikivie, P. Experimental tests of the "invisible" axion. {\em Phys. Rev. Lett.} {\bf 1983}, {\em 51}, 1415--1417.

\bibitem[Wilczek(1987)]{Ax3}
Wilczek, F. Two applications of axion electrodynamics. {\em Phys. Rev. Lett.} {\bf 1987}, {\em 58}, 1799--1802.

\bibitem[Bertone(2005)]{ADM1}
Bertone, G.; Hooper, D.; Silk, J. Particle Dark Matter: Evidence, Candidates and Constraints. {\em Physics Reports} {\bf 2005}, {\em 405}, 279--390.

\bibitem[Duffy(2009)]{ADM11}
Duffy, L.D.; van Bibber, K. Axions as dark matter particles. {\em New J. Phys.} {\bf 2009}, {\em 11}, 105008.

\bibitem[Khlopov(2012)]{ADM2}
Khlopov, M. \textit{Fundamentals of Cosmic Particle Physics}; CISP-Springer: Cambridge, UK, 2012.

\bibitem[Del Popolo(2014)]{ADM22}
Del Popolo, A. Nonbaryonic dark matter in cosmology. {\em Int. J. Mod. Phys. D} {\bf 2014}, {\em 23}, 1430005.

\bibitem[Marsh(2016)]{ADM3}
Marsh, D.J.E. Axion cosmology. {\em Physics Reports} {\bf 2016}, {\em 643}, 1--79.

\bibitem[LIGO(2017)]{GRB170817}
Abbott, B.P. et.al [LIGO Scientific Collaboration, Virgo Collaboration, Fermi Gamma-Ray Burst Monitor, INTEGRAL]. Gravitational Waves and Gamma-rays from a Binary Neutron Star Merger:
GW170817 and GRB 170817A. {\em Astrophys. J. Lett.} {\bf 2017}, {\em 848}, L13.


\end{thebibliography}
\end{document}